\documentclass{jabad}

\usepackage{graphicx}

\begin{document}

\title[Counting in the BP Landscape]{Counting states in the\\
  Bousso-Polchinski Landscape}

\newcommand{\shortauthor}{{C.~Asensio and A.~Segui}}

\author{C\'esar Asensio\footnote{casencha@unizar.es} and
  Antonio Segu\'{\i}\footnote{segui@unizar.es}
}

\begin{center}
 {Departamento de F\'{\i}sica Te\'{o}rica, Facultad de Ciencias,\\
  Universidad de Zaragoza, 50009 Zaragoza, Spain}
\end{center}

\begin{abstract}
  Starting from an exact counting of small and positive cosmological
  constant states in the Bousso-Polchinski Landscape we recover a
  well-known approximate formula and a systematic method of
  improvement by means of the Poisson summation formula.  %
  This is a contribution to the special Volume published by the
  University of Zaragoza in honor of Julio Abad Anto\~nanzas.
  En memoria de nuestro amigo, compa\~nero y maestro Julio.
\end{abstract}

\KeysAndCodes{Bousso-Polchinski Landscape, Cosmological Constant}
{04.60.-m,11.25.-w}

\section{Introduction}

\newcommand{\dif}{\,{\rm d}}
\newcommand{\vol}{\mathop{\rm vol}}

One of the recent proposals to solve the cosmological constant problem
in cosmology is provided by string theory.  By dimensional reduction
from M-theory to 3+1 dimensions, vacua of the effective theory are
classified by means of a big number of quantized fluxes leading to an
enormous amount of metastable vacua, the Bousso-Polchinski (BP)
Landscape \cite{BP}.  The cosmological constant problem, namely the
smallness of the observed vacuum energy density in the universe
\cite{WW2,B-CC}, can be solved by the presence in this model of a huge
number of states of very small, positive cosmological constant,
together with a dynamical mechanism given by eternal inflation
\cite{EtInf} which allows the system to visit all the vacua.  An
anthropic selection is then advocated to explain the smallness of the
observed cosmological constant \cite{Anthr,WW}.

In order to quantify this selection a counting of accesible states in
the Landscape is needed.  The simplest one is the Bousso-Polchinski
count, which computes the volume of a spherical shell of small
thickness in flux space and divides it by the volume of a cell.  We
will now briefly review this argument (see \cite{BP}).

A vacuum of the BP Landscape is a node in a $J$-dimensional lattice
${\cal L}$ generated by $J$ charges $q_1,\cdots,q_J$ determined by the
sizes of the three-cycles in the compactification manifold.  The
lattice ${\cal L}$ is
\begin{equation}
  \label{eq:c1}
  \mathcal{L} = \bigl\{(n_1q_1,\cdots,n_Jq_J)\in\Rset^J\colon
  n_1,\cdots,n_J\in\Zset\bigr\}\,.
\end{equation}
The $j$-th coordinate of a point in the lattice is an integer multiple
of the charge $q_j$, and therefore a vacuum is characterized by the
integer $J$-tuple $n=(n_1,\cdots,n_J)$.

A \emph{fundamental cell} (also called \emph{Voronoi
  cell}\footnote{Also called Wigner-Seitz cell in solid state physics,
  the Voronoi cell of a point $P$ in a discrete set $S$ of a metric
  space $M$ is the set of points of $M$ which are closer to $P$ than to
  any other point of $S$.}) $Q_n$ around a node $n$ in a lattice
${\cal L}$ is the subset of $\Rset^J$ which contains the points which are
closer to $n$ than to any other node of ${\cal L}$.  Thanks to the
discrete translational symmetry of our lattice (\ref{eq:c1}), all
fundamental cells in ${\cal L}$ are translates of the fundamental cell
around the origin $Q_O\equiv Q$, which we can parametrize in Cartesian
coordinates as a product of symmetric intervals
\begin{equation}
  \label{eq:c2}
  Q = \prod_{j=1}^J\Bigl[-\frac{q_j}{2},\frac{q_j}{2}\Bigr]\,.
\end{equation}
The cosmological constant of vacuum $n$ in the BP model is\footnote{We
  use reduced Planck units in which $8\pi G = \hbar = c = 1$.}
\begin{equation}
  \label{eq:c3}
  \Lambda(n) = \Lambda_0 + \frac{1}{2}\sum_{j=1}^J n_j^2q_j^2\,.
\end{equation}
In (\ref{eq:c3}), $\Lambda_0$ is an \emph{a priori} cosmological
constant or order $-1$.  Each value of $\Lambda>\Lambda_0$ defines a
spherical ball on the $J$-dimensional flux space of radius $R_\Lambda
= \sqrt{2(\Lambda-\Lambda_0)}$.  We call this ball ${\cal
  B}^{J}(\Lambda)$.  We take small values of the charges $q_j$
(natural values expected by BP are of order $\frac{1}{6}$) in such a
way that the ball can contain a huge number of fundamental cells.

The number of states in the Weinberg Window, that is the range of
values of the cosmological constant allowing the formation of
structures (like galaxies) needed for the formation of life as we know
it \cite{WW}, is obtained by computing the volume of a thin spherical
shell in flux space (the realization in the BP Landscape of the
Weinberg Window) divided by the volume of a cell in the lattice:
\begin{equation}
  \label{eq:c4}
  \begin{split}
    \mathcal{N}_{\text{WW}}
    &= \frac{\vol {\cal B}^J(\Lambda_{\text{WW}}) - \vol {\cal B}^J(0)}{\vol Q}
    \approx \frac{1}{\vol Q}
    \left.
      \frac{\dif}{\dif\Lambda}\Bigl(\vol{\cal B}^J(\Lambda)\Bigr)
    \right|_{\Lambda=0}\Lambda_{\text{WW}} \\
    &= \frac{1}{\vol Q}
    \left.
      \frac{\dif}{\dif\Lambda}
      \Bigl(\frac{R_\Lambda^J}{J}\vol{S}^{J-1}\Bigr)
    \right|_{\Lambda=0}\Lambda_{\text{WW}}
    = \vol{S}^{J-1}\frac{R_0^{J-2}\Lambda_{\text{WW}}}{\vol Q}\,,
  \end{split}
\end{equation}
where $R_0 = R_{\Lambda = 0} = \sqrt{2|\Lambda_0|}$ (we will call it
henceforth $R$), and the volume of the $J-1$ dimensional sphere is
\begin{equation}
  \label{eq:c5}
  \vol S^{J-1} =
  \frac{2\pi^{\frac{J}{2}}}{\Gamma\bigl(\frac{J}{2}\bigr)}\,.
\end{equation}
This method can be naively expected to yield a good estimate when the
linear dimensions of the cell are small when compared to the thickness
of the shell; but this condition is not satisfied in the BP Landscape.
Nevertheless, the result of this counting formula is very good when
compared to actual numerical experiments.  In the following, we will
re-derive the BP count, systematic improvements and a condition of
validity.

Our proposal is based on the following kinds of states one may
encounter near the null cosmological constant surface in flux space:
\begin{itemize}
\item \textbf{Boundary} (or \textbf{penultimate} after Bousso and Yang
  \cite{BY}) are those states in which a Brown-Teitelboim
  \cite{BT-1,BT-2} decay chain can end before jumping into the
  negative cosmological constant sea.  So we define a boundary state
  as one having
  \begin{itemize}
  \item [(1)] positive cosmological constant, and
  \item [(2)] at least one neighbor of negative cosmological constant.
  \end{itemize}
\item \textbf{Secant} states have the property that their Voronoi
  cells in flux space have non-empty intersection with the null
  cosmological constant surface in flux space.  Note that a secant
  state may have negative cosmological constant.
\end{itemize}
These two categories are not equivalent; a boundary state may not be
secant if it is far enough from the null cosmological constant
surface, and a secant state may not be boundary if it has negative
cosmological constant.  So we are interested mainly in the states
which are both secant \textbf{and} boundary, because all the states in
the Weinberg Window are in this category.

Our strategy will be as follows.  We will count the states in the
Weinberg Window using the secant states instead of the boundary
states.  We approximate the exact count using the Poisson summation
formula in section \ref{sec:N_S}.  The same technique is used to
obtain a distribution of values of the cosmological constant and the
number of states in the Weinberg Window in section \ref{sec:N_WW}.
Our results yield the BP count as the lowest order approximation as
well as systematic improvements.  In section \ref{sec:N_B} we sketch
the difficulties encountered while extending the method to the
boundary states.  Finally, we summarize the conclusions in section
\ref{sec:conc}.

This work is a continuation of the counting method introduced in
\cite{AS-RHM}.

\section{Counting secant states}
\label{sec:N_S}

\begin{figure}[htbp]
  \centering
  \includegraphics{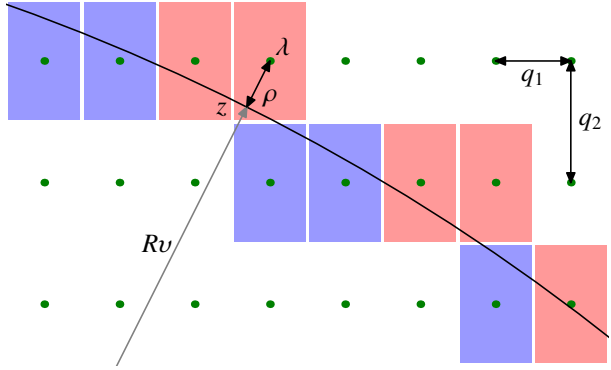}
  \caption{Secant states are shown in a $J=2$ BP model along with the
    definition of the $\rho$ and $\upsilon$ parameters.}
  \label{fig:1}
\end{figure}

If $\lambda=(n_1q_q,\cdots,n_Jq_J)$ is a secant state of the lattice
$\mathcal{L}$, the $\Lambda=0$ sphere intersects its Voronoi cell.
The line which links the origin with $\lambda$ hits the sphere in a
single point, $z$.  The directions of $\lambda$ and $z$ are the same,
$\upsilon\in S^{J-1}$, and their norms are $\|z\| = R$ and
$\|\lambda\| = R + \rho$.  The parameter $\rho$ is simply the distance
between the sphere and $\lambda$ (see figure \ref{fig:1}), and has a
close relation with the cosmological constant:
\begin{equation}
  \label{eq:c6}
  \rho = \sqrt{2(\Lambda - \Lambda_0)} - \sqrt{2|\Lambda_0|}\,.
\end{equation}
In particular, $\Lambda$ is positive if and only if $\rho$ is
positive, and $\Lambda=0$ if and only if $\rho=0$.

Note that the pair $(\rho,\upsilon)$ determines the tangent hyperplane
on the $\Lambda=0$ sphere at $z$, and therefore can be identified with
this hyperplane $h=(\rho,\upsilon)$.  Note that $h$ intersects the
Voronoi cell of $\lambda$, but $z$ is not necessarily inside the cell.
The set of possible $z$ points, each corresponding to a secant
hyperplane, can be specified by computing the maximum value which
$\rho$ can have for a fixed direction $\upsilon$.  We will call this
quantity $\rho_{\text{max}}=\sigma(\upsilon)$, which is given by the
distance of the most distant secant hyperplane
\begin{equation}
  \label{eq:c7}
  \upsilon\cdot(x - c_\upsilon) = 0\,,
\end{equation}
where $c_\upsilon=\frac{1}{2}(s_1q_1,\cdots,s_Jq_J)$ is the unique
corner out of $2^J$ which belongs to the same $J$-quadrant as
$\upsilon$, and the $s_j$ are signs $\pm1$, indeed the same signs of
the components of $\upsilon$.  We find 
\begin{equation}
  \label{eq:c8}
  \sigma(\upsilon) = \upsilon\cdot c_\upsilon
  = \frac{1}{2}\sum^J_{j=1} q_j s_j^2 |\upsilon_j| =
  \frac{1}{2}\sum^J_{j=1}q_j|\upsilon_j| = \frac{1}{2}q\cdot|\upsilon|\,.
\end{equation}
In (\ref{eq:c8}), $|\upsilon|=(|\upsilon_1|,\cdots,|\upsilon_J|)$.
Note that the function $\sigma(\upsilon)$ defines a surface in flux
space which contains the possible $z$ points, equivalently, the
possible secant hyperplanes.  We call the interior of this surface the
\emph{(secant) hyperplane space} associated to the cell, $H_Q$, and a
hyperplane $h=(\rho,\upsilon)$ is in $H_Q$ if and only if
$\rho\le\sigma(\upsilon)$.

\begin{figure}[htbp]
  \centering
  \includegraphics{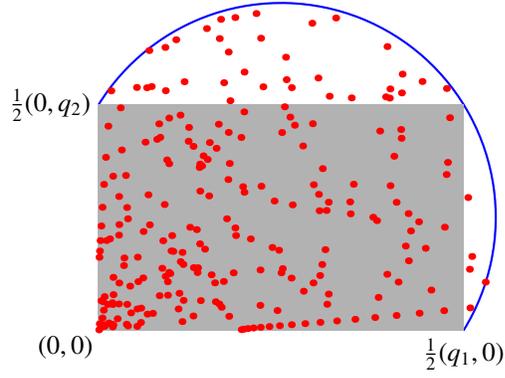}
  \caption{Secant hyperplanes of positive $\rho$ in a $J=2$ BP
    landscape.  Each dot is a $z$ point of a secant state.  The shaded
    region is the first quadrant of the cell $Q$, and all dots lie
    inside the blue contour, which is the (first quadrant) boundary of
    hyperplane space $H_Q$.}
  \label{fig:2}
\end{figure}

If we take all secant states of a given lattice and we gather all
their cells into one, the $z$ points seem to be randomly distributed
inside $H_Q$ (see figure \ref{fig:2}).  So our suggestion is to
provide an explicit form for the probability measure which governs
this random distribution.  We call the choosing of a particular
measure $\dif P(\rho,\upsilon)$ in $H_Q$ a \emph{random hyperplane
  model} (RHM).

In our previous work \cite{AS-RHM} we chose the uniform probability
measure, and we will justify this assumption as an approximation in
section \ref{sec:N_WW}.

\begin{figure}[htbp]
  \centering
  \includegraphics[width=\textwidth]{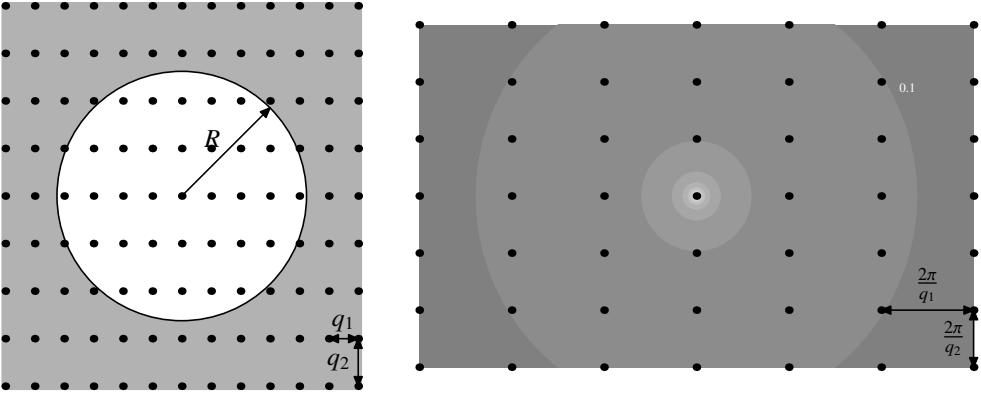}
  \caption{$J=2$ lattice with a typical scale $R$ (left).  Dual
    lattice (right).  The shaded circles show decreasing amplitude of
    the envelope of the Bessel function $J_0(R\|\xi\|)$, which is the
    Fourier transform of the normalized indicator function of the
    circle of radius $R$ in the former lattice.  The 0.1 label signals
    the distance in dual space where the amplitude at the origin is
    reduced to 1/10.}
  \label{fig:3}
\end{figure}

We start using an exact expression for the number of secant states of
positive cosmological constant.  We call this number $\mathcal{N}_S$.
For each secant state $\lambda$, we denote by $\rho(\lambda)$ and
$\sigma(\lambda)$ the parameters of its associated secant hyperplane,
being simply
\begin{equation}
  \label{eq:c9}
  \rho(\lambda) = \|\lambda\|-R\,,\quad
  \sigma(\lambda) = \frac{1}{2}q\cdot\frac{|\lambda|}{\|\lambda\|}\,.
\end{equation}
Using the restriction $\rho\le\sigma(\upsilon)$ and the indicator
function $\chi_I(x)$ of an interval $I\subset\Rset$ (which is 1 if
$x\in I$ and 0 otherwise), we can write
\begin{equation}
  \label{eq:c10}
  \mathcal{N}_S =
  \sum_{\lambda\in\mathcal{L}}\chi_{[0,\sigma(\lambda)]}[\rho(\lambda)]\,.
\end{equation}
Direct use of (\ref{eq:c10}) is unfeasible; we would have to compile
all secant states to count them.  But it has the form
\begin{equation}
  \label{eq:c11}
  \sum_{\lambda\in\mathcal{L}}f(\lambda)
  \quad\text{with}\quad
  f(x) = \chi_{[0,\sigma(x)]}[\rho(x)]
 \quad(x\in\Rset^J)\,,
\end{equation}
and therefore we can obtain an alternative representation using the
Poisson summation formula:
\begin{equation}
  \label{eq:c12}
  \sum_{\lambda\in\mathcal{L}}f(\lambda) =
  \frac{1}{\vol Q}
  \sum_{\kappa\in\mathcal{L}^*}\hat f(\kappa)
\end{equation}
where we use the Fourier transform of $f(x)$
\begin{equation}
  \label{eq:c13}
  \hat f(\xi) = \int_{\Rset^J}f(x)e^{-i\xi\cdot x}\dif^Jx\,,
\end{equation}
and the dual lattice of $\mathcal{L}$
\begin{equation}
  \label{eq:c14}
  \mathcal{L}^* =
  \biggl\{\biggl(m_1\frac{2\pi}{q_1},\cdots,m_J\frac{2\pi}{q_J}\biggl)
  \in\Rset^J\colon m_1,\cdots,m_J\in\Zset\biggr\}\,,
\end{equation}
determined by the condition $e^{i\lambda\cdot\kappa} = 1$.  In
fig.~\ref{fig:3} a $J=2$ lattice and its dual are shown.

In order to use eq.~(\ref{eq:c12}) we need to compute
\begin{equation}
  \label{eq:c15}
  \hat f(\xi) = \int_{\Rset^J}\chi_{[0,\sigma(x)]}[\rho(x)]
  e^{-i\xi\cdot x}\dif^Jx\,.
\end{equation}
Switching to spherical variables and then using $\rho$ as integration
variable, we obtain
\begin{equation}
  \label{eq:c16}
  \hat f(\xi) = \int_{S^{J-1}}\dif\Omega_{J-1}(\upsilon)\,
  e^{-iR(\xi\cdot\upsilon)}
  \int^{\sigma(\upsilon)}_0 (\rho+R)^{J-1}
  e^{-i\rho(\xi\cdot\upsilon)}\dif\rho\,.
\end{equation}
Exact evaluation of this integral is difficult, as might be expected.
The lowest order approximation in (\ref{eq:c12}) yields the following
formula for the number of positive $\Lambda$ secant states:
\begin{equation}
  \label{eq:c17}
  \mathcal{N}_S = \frac{\hat f(0)}{\vol Q}
  = \frac{1}{\vol Q}
  \int_{S^{J-1}}\dif\Omega_{J-1}(\upsilon)\,
  \int^{\sigma(\upsilon)}_0 (\rho+R)^{J-1} \dif\rho\,.
\end{equation}
We can expect this to be an accurate approximation from the following
qualitative argument.  The function $f(x)$ is supported at a compact
domain whose extent is proportional to some average charge
$\overline{q}$ times a $J$-dependent constant.  The Fourier transform
$\hat f(\xi)$ will be then concentrated at a domain of dimensions
proportional to $1/\overline{q}$.  But the spacings between
neighboring nodes of the dual lattice are the inverses of the charges,
and thus only a few of these nodes will enter the domain with
significant amplitude.

In this approximation, we can rewrite eq.~(\ref{eq:c17}) as
\begin{equation}
  \label{eq:rhm-1}
  \frac{1}{\mathcal{N}_S\vol Q}
  \int_{S^{J-1}}\dif\Omega_{J-1}(\upsilon)\,
  \int^{\sigma(\upsilon)}_0 (\rho+R)^{J-1} \dif\rho = 1\,,
\end{equation}
which asserts that the following probability measure is normalized:
\begin{equation}
  \label{eq:rhm-2}
  \dif P(\rho,\upsilon) = \chi_{[0,\sigma(\upsilon)]}(\rho)
  \frac{(\rho+R)^{J-1} \dif\rho\dif\Omega_{J-1}(\upsilon)}{\mathcal{N}_S\vol Q}\,.
\end{equation}
The $\chi$ function simply restricts the range of integration to the
secant hyperplane set $H_Q$.  The previous measure can be expressed
using the $z$ points mentioned above translated to a single cell $Q$,
that is, considering $z\mod{\mathcal{L}}$, the measure
(\ref{eq:rhm-2}) is
\begin{equation}
  \label{eq:rhm-3}
  \dif P(z) = \chi_{H_Q}(z)\frac{\dif^J z}{\mathcal{N}_S\vol Q}\,,
\end{equation}
that is, the uniform probability in $H_Q$.  So the random hyperplane
model obtained to the lowest order in the Poisson summation formula is
simply the uniform probability measure over $H_Q$. This is the choice
we made in \cite{AS-RHM}.

Integrating out the $\rho$ variable in (\ref{eq:c17}) we find
\begin{equation}
  \label{eq:c18}
  \begin{split}
    \mathcal{N}_S &= \frac{1}{J\vol Q} \int_{S^{J-1}}
    \bigl[(R+\sigma(\upsilon)\bigr)^J - R^J\bigr]
    \dif\Omega_{J-1}(\upsilon) \\
    &= \frac{1}{J\vol Q}
    \sum^J_{k=1} \binom{J}{k}R^{J-k}
    \int_{S^{J-1}}\sigma(\upsilon)^k
    \dif\Omega_{J-1}(\upsilon)
    \,.
  \end{split}
\end{equation}
The first term in this expansion can be computed exactly (this is done
in \cite{AS-RHM}):
\begin{equation}
  \label{eq:c19}
  \int_{S^{J-1}}\sigma(\upsilon)
  \dif\Omega_{J-1}(\upsilon) =
  \frac{\vol S^{J-2}}{J-1}\sum^J_{j=1}q_j\,,
\end{equation}
so that we recover the formula obtained in \cite{AS-RHM} with a
completely different approach:
\begin{equation}
  \label{eq:c20}
  \mathcal{N}_S = \frac{\vol S^{J-2}}{J-1}\,\frac{R^{J-1}}{\vol Q}
  \sum^J_{j=1}q_j\,.
\end{equation}
Note the absence of a factor 2 here because of the counting of
\emph{positive} $\Lambda$ states only.

The validity condition of formula (\ref{eq:c20}) is
\begin{equation}
  \label{eq:c21}
  \sigma(\upsilon) \le \sigma_{\text{max}}
  = \frac{1}{2}\sqrt{\sum^J_{j=1}q_j^2}
  \lll R
  \,.
\end{equation}
Using a mean square charge $\tilde{q}^2 =
\frac{1}{J}\sum^J_{j=1}q_j^2$, condition~(\ref{eq:c21}) can be
rewritten as\footnote{Incidentally, the adimensional parameter
  occurring in (\ref{eq:c22}) resembles the so-called \emph{planar
    limit} in field theory, in which the number $N$ characterizing the
  gauge group tends to infinity and the Yang-Mills coupling constant
  $g_{\text{YM}}$ vanishes with the product $Ng_{\text{YM}}^2$ (the
  t'Hooft coupling) held fixed.}
\begin{equation}
  \label{eq:c22}
  J\frac{\tilde{q}^2}{|\Lambda_0|} \lll 8\,.
\end{equation}

\section{Number of states in the Weinberg Window}
\label{sec:N_WW}

We can follow the same steps for the number of states of very low
cosmological constant.  If $\Lambda_{\text{WW}}$ is a very small
number characterizing the anthropic range (which comprises the
experimental value $\Lambda_\text{exp}\approx10^{-123}$), we will call
$\rho_a$ the corresponding value of the $\rho$ parameter computed
using eq.~(\ref{eq:c6}), that is
\begin{equation}
  \label{eq:c23}
  \rho_a = \sqrt{2(\Lambda_{\text{WW}} - \Lambda_0)} -
  \sqrt{2|\Lambda_0|}
  \approx \frac{\Lambda_{\text{WW}}}{R}\,.
\end{equation}
The number of states in the BP lattice having
$0\le\rho(\lambda)\le\tau$ for fixed $\tau$ is
\begin{equation}
  \label{eq:c24}
  \Omega(\tau) =
  \sum_{\lambda\in\mathcal{L}}
  \chi_{[0,\sigma(\lambda)]\cap[0,\tau]}[\rho(\lambda)]\,.
\end{equation}
Note that $\Omega(\tau) = \mathcal{N}_S$ if $\tau >
\sigma_{\text{max}}$, so that interpreting each secant state as a
random equiprobable event we obtain the general formula
\begin{equation}
  \label{eq:c25}
  \Omega(\tau) = \mathcal{N}_S P(0 \le \rho \le \tau)\,.
\end{equation}
On the other hand, if $\tau < \frac{1}{2}\min_j\{q_j\} =
\sigma_{\text{max}}$, then $[0,\sigma(\lambda)]\cap[0,\tau] =
[0,\tau]$ and we have $f(x)=\chi_{[0,\tau]}[\rho(x)]$, which
simplifies the Fourier transform (\ref{eq:c16}) to
\begin{equation}
  \label{eq:c26}
  \hat f(\xi) = \int_{S^{J-1}}\dif\Omega_{J-1}(\upsilon)\,
  e^{-iR(\xi\cdot\upsilon)}
  \int^{\tau}_0 (\rho+R)^{J-1}
  e^{-i\rho(\xi\cdot\upsilon)}\dif\rho\,.
\end{equation}
For $\tau$ as small as $\rho_a$, a first order Taylor expansion around
$\tau=0$ yields
\begin{equation}
  \label{eq:c27}
  \hat f(\xi) = R^{J-1}\tau
  \int_{S^{J-1}}\dif\Omega_{J-1}(\upsilon)\,
  e^{-iR(\xi\cdot\upsilon)}
  \,.
\end{equation}
Taking only the first term in the Poisson summation formula we obtain
\begin{equation}
  \label{eq:c28}
  \Omega(\tau) \approx \frac{\hat f(0)}{\vol Q}
  = \frac{R^{J-1}}{\vol Q}\,\vol S^{J-1}\,\tau\,,
\end{equation}
which is the number of states in the Weinberg Window once we
specialize $\tau=\rho_a=\frac{\Lambda_{\text{WW}}}{R}$ in equation
(\ref{eq:c28}):
\begin{equation}
  \label{eq:c29}
  \mathcal{N}_{\text{WW}} = \frac{R^{J-2}}{\vol Q}\,\vol S^{J-1}\,
  \Lambda_{\text{WW}}\,.
\end{equation}
This formula is exactly the Bousso-Polchinski count given in \cite{BP}
and rederived in \cite{AS-RHM} using the RHM.  

It turns out that the Fourier transform (\ref{eq:c27}) can be given in
closed form, allowing systematic improvements to be added to
(\ref{eq:c29}).  To show this, we rewrite (\ref{eq:c27}) as
\begin{equation}
  \label{eq:c30}
  \hat f(\xi) = R^{J-1}\tau
  \int_{S^{J-1}}\dif\Omega_{J-1}(\upsilon)\,
  e^{-iR(\xi\cdot\upsilon)} =
  \tau\frac{\partial}{\partial R}
  \int_{\Rset^J}\theta(R^2 - \|x\|^2)\,
  e^{-i\xi\cdot x}\dif^Jx
  \,.
\end{equation}
Now, we use the following integral representation for the $\theta$ function:
\begin{equation}
  \label{eq:c31}
  \theta(t) = \frac{1}{2\pi i}\int_{\gamma}\frac{e^{st}}{s}\dif s\,,
\end{equation}
where $\gamma$ is a contour running from $c-i\infty$ to $c+i\infty$
for real $c$ located to the right side of all singularities of the
integrand.  We have
\begin{equation}
  \label{eq:c32}
  \begin{split}
    \hat f(\xi) &= 
    \frac{\tau}{2\pi i}\int_{\gamma}\frac{\dif s}{s}
    \frac{\partial}{\partial R}
    \int_{\Rset^J}e^{s(R^2 - \|x\|^2) - i\xi\cdot x}\dif^J x \\
    &= \frac{2R\tau}{2\pi i}\int_{\gamma}e^{sR^2}\dif s
    \prod^J_{j=1}\int_{\Rset} e^{-sx_j^2 - i\xi_jx_j}\dif x_j \\
    &= \frac{2R\tau}{2\pi i}\int_{\gamma}e^{sR^2}\dif s
    \prod^J_{j=1}\sqrt{\frac{\pi}{s}}\, e^{-\frac{\xi_j^2}{4s}} \\
    &= \frac{2R\tau\pi^{\frac{J}{2}}}{2\pi i}\int_{\gamma}
    e^{sR^2 - \frac{\|\xi\|^2}{4s}}\frac{\dif s}{s^{\frac{J}{2}}} \\
    &= 2R\tau\pi^{\frac{J}{2}}
    \biggl(\frac{2R}{\|\xi\|}\biggr)^{\frac{J}{2}-1}
    J_{\frac{J}{2}-1}(R\|\xi\|)
    \,.
  \end{split}
\end{equation}
The last step is an inverse Laplace transform resulting in a Bessel
function.  Let us write this result in a slightly more convenient
form, using $\nu=\frac{J}{2}-1$:
\begin{equation}
  \label{eq:c33}
  \hat f(\xi) = R\tau \, R^{J-2}
  \frac{2\pi^{\frac{J}{2}}}{\Gamma\bigl(\frac{J}{2}\bigr)}
  L_\nu(R\|\xi\|)
  \,,\quad\text{with}\quad
  L_\nu(t) = 
  \Gamma(\nu+1) 2^\nu
  \frac{J_{\nu}(t)}{t^\nu}
  \,.
\end{equation}
The function $L_\nu(t)$ satisfies $L_\nu(0)=1$ and has a gaussian
shape near $t=0$, $L_\nu(t)\approx e^{-\frac{t^2}{4(\nu+1)}}$,
followed by a regime of damped oscillations with an envelope
proportional to $t^{-\nu-\frac{1}{2}}$.  We can use this expression to
write the first correction to (\ref{eq:c29}) by choosing the $2J$
neighbors of the origin in the dual lattice.  For these,
$\|\xi\|=\frac{2\pi}{q_j}$ for each axis, so we have
\begin{equation}
  \label{eq:c34}
  \mathcal{N}_{\text{WW}} = \Lambda_{\text{WW}}
  \frac{R^{J-2}}{\vol Q}\,\vol S^{J-1}\,
  \biggl[1 + 2\sum^J_{j=1}L_{\nu}(2\pi R/q_j)\biggr]
  \,.
\end{equation}
The first correction will be small when compared to (\ref{eq:c29})
if
\begin{equation}
  \label{eq:c35}
  2\biggl|
    \sum^J_{j=1}L_{\nu}(2\pi R/q_j)
  \biggr| \ll 1\,.
\end{equation}
Condition (\ref{eq:c35}) will be satisfied as long as the $J$ numbers
$2\pi\frac{R}{q_j}$ are scattered along regions of different sign of
the Bessel function, allowing a huge cancellation in the sum.  The
worst case will occur when all charges are equal (let $q$ be their
common value), and then we must demand that $2\pi\frac{R}{q}$ is very
far from the gaussian regime of the function $L_\nu(t)$, that is,
\begin{equation}
  \label{eq:c36}
  2\pi\frac{R}{q} \gg k\sqrt{J}
  \quad\Rightarrow\quad
  J\frac{q^2}{|\Lambda_0|} \ll \frac{8\pi^2}{k}\,,
\end{equation}
where the number $k$ is $J$-dependent.  For $J=2$ ($\nu=0$), taking
$10^{-2}$ as a reference amplitude for a relative error of $1\%$, we
need a value of $k\sqrt{2}$ large enough to be in the asymptotic
regime of the Bessel function, so we have
\begin{equation}
  \label{eq:c37}
  4\sqrt{\frac{2}{\pi k\sqrt{2}}} = 10^{-2}
  \quad\Rightarrow\quad
  k \approx 72025.3\,,
\end{equation}
which implies $q\ll 0.01655R$ from (\ref{eq:c36}), but for $J=22$
($\nu=10$) the value $10^{-2}$ is in the gaussian regime, and we have
\begin{equation}
  \label{eq:c38}
  44e^{-\frac{k^2}{2}} = 10^{-2}
  \quad\Rightarrow\quad
  k \approx 4.09618\,,
\end{equation}
which implies $q\ll 0.66188R$.  We see that the requirements in the
charges are not too restrictive so that the Bousso-Polchinki count
(\ref{eq:c29}) can be reasonably accurate.

We can also expect the subsequent corrections to be small when
compared to the first one because of the cancellations in the sum of
Bessel functions.

\section{Replacing secant by boundary states}
\label{sec:N_B}

If we want to use boundary states to count the low-$\Lambda$ states,
we must find a condition satisfied by them which is analogous to
$\rho<\sigma(\upsilon)$ for the secant states.  A boundary state has
positive $\Lambda$ but some neighbors (possibly only one) of negative
$\Lambda$, which means that the $\Lambda=0$ sphere must cut the
segment joining both states.  The segments joining a boundary state
with its $2J$ neighbors constitute the skeleton of a cell which have
these neighbors as vertices and a face for each ``$J$-quadrant''
determined by the $J$ vertices in it.  For $J=2$, such a cell is
simply a rhombus with the neighbors at its corners, and for $J=3$ the
cell is an octahedron.  We will call this cell a \emph{$J$-rhombus}.

These cells are not disjoint, and do not cover the whole flux space,
so we cannot use them for tessellate flux space in order to count
states naively.  But we can reformulate the condition of a boundary
state as being secant with respect to this cell instead of its Voronoi
cell.  Note that the $\Lambda=0$ surface intersecting a $J$-rhombus
will have its $\rho$ and $\upsilon$ parameters defined, so the
boundary condition is again $\rho<\sigma(\upsilon)$ for a
different\footnote{We will call also $\sigma$ to this function in this
  section; this should not lead the reader to confusion.}  $\sigma$
interpreted as the boundary of the space of hyperplanes associated to
boundary states.

\begin{figure}[htbp]
  \centering
  \includegraphics{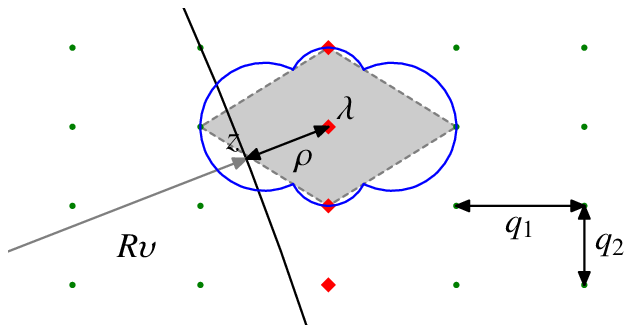}
  \caption{Boundary states (diamonds) in a $J=2$ BP lanscape.  The
    shaded region is the $2$-rhombus. The boundary of the boundary
    hyperplane space is also displayed.  Note that the $z$ point is
    slightly outside the rhombic cell.}
  \label{fig:4}
\end{figure}

Again, $\sigma(\upsilon)$ is the maximum distance that a hyperplane
associated to a boundary state (``boundary hyperplane'' henceforth)
can reach in a given direction $\upsilon$.  The distance will be
maximum if the hyperplane contains at least one vertex of the
$J$-rhombus.  If we assume the center of the $J$-rhombus to be at the
origin, the vertices are the points $s_iq_ie_i$, where $s_i=\pm1$ and
$e_i$ is the unit vector of the $i^{\text{th}}$ axis.  The equation of
a maximum distance boundary hyperplane is
\begin{equation}
  \label{eq:c39}
  v\cdot(x-s_iq_ie_i) = 0\,,
\end{equation}
so that the maximum distance is $\sigma(\upsilon)=|\upsilon_i|q_i$.
The value of $i$ and $s_i$ are chosen in the following way.  Let $u$
be the normal unit vector to the face of the $J$-rhombus in the first
(all positive components) $J$-quadrant, $S^{J-1}_+$:
\begin{equation}
  \label{eq:c40}
  u = \frac{(q_1^{-1},\cdots,q_J^{-1})}{\sum^J_{j=1}q_j^{-2}}\,.
\end{equation}
Point $u$ decomposes $S^{J-1}_+$ in the following $J$ regions:
\begin{equation}
  \label{eq:c41}
  \Xi_i = \Bigl\{\upsilon\in S^{J-1}_+\colon \upsilon_i \ge u_i,
  \ {}
  \upsilon_j \le u_j\quad\text{for $j\ne i$}\Bigr\}\,.
\end{equation}
Now, let us move $\upsilon$ to $S^{J-1}_+$ by removing the signs:
$|\upsilon| = (|\upsilon_1|,\cdots,|\upsilon_J|)$.  Choose the label
$i$ for the set $\Xi_i$ which $\upsilon$ belongs to, and choose the
sign $s_i$ to be that of $\upsilon_i$.  Thus, we find
\begin{equation}
  \label{eq:c42}
  \sigma(\upsilon) = \sum^J_{i=1}q_i|\upsilon_i|\chi_{\Xi_i}(|\upsilon|)\,.
\end{equation}
For $J=2$ we have $u\propto(\frac{1}{q_1},\frac{1}{q_2})$ and this
determines an angle $\theta_0 = \tan^{-1}\frac{q_1}{q_2}$.  $\Xi_1$
comprises the directions $\upsilon=(\cos\theta,\sin\theta)$ for which
$\theta\le\theta_0$ and $\Xi_2$ the converse $\theta\ge\theta_0$.  An
example of this boundary is shown in figure \ref{fig:4}.

Once the $\sigma$ function has been found, we can write a formula
analogous to (\ref{eq:c18}) for the number of boundary states, whose
first term is
\begin{equation}
  \label{eq:c43}
  \mathcal{N}_B = \frac{R^{J-1}}{\vol Q}
  \int_{S^{J-1}}\sigma(\upsilon)
  \dif\Omega_{J-1}(\upsilon)\,.
\end{equation}
The validity condition now is
\begin{equation}
  \label{eq:c44}
  \sigma(\upsilon) \le \sigma_{\text{max}}
  = \max_{1\le j\le J}\{q_j\}
  \lll R
  \,.
\end{equation}

But we do not need to evaluate formula~(\ref{eq:c43}) if we only want
to count states in the Weinberg Window.  Note that equation
(\ref{eq:c24}) does not involve $\sigma$ because of the smallness of
the shell width, and therefore the number $\mathcal{N}_{\text{WW}}$
computed in section \ref{sec:N_WW} remains unchanged.  Only a
difference must be stressed in this regard:  the shell width must be
smaller than the minimum distance $\sigma_{\text{min}}$, which now
depends on $J$.  It is simply the distance to the cell center of one
of its faces.  The face at $S^{J-1}_+$ has equation
$\sum^J_{j=1}\frac{x_j}{q_j} = 1$ or $u\cdot x = \sigma_{\text{min}}$,
so we have the condition
\begin{equation}
  \label{eq:45}
  \sigma_{\text{min}} =
  \biggl[\sum^J_{j=1}q_j^{-2}\biggr]^{-\frac{1}{2}}
  = \frac{\widetilde{q}}{\sqrt{J}}
  \gg \rho_a = \frac{\Lambda_{\text{WW}}}{R}\,,
\end{equation}
where $\widetilde{q}$ is a ``square-harmonic'' average of the charges.
Note that this condition is met for natural values of the charges and
dimension $J$, so that considering secant or boundary states may
change the probability distribution of $\Lambda$ in equation
(\ref{eq:c25}) but it has no effect on the number of states in the
Weinberg Window.

\section{Conclusion}
\label{sec:conc}

We have developed a general method for counting low-lying states in
the Bousso-Polchinki Landscape with the help of the Poisson summation
formula without any statistical validation of the assumptions made.
This approach provides a firm foundation of the random hyperplane
model previously used by the authors.  It also allows us to derive
validity conditions as well as systematic improvements, and can be
used in different problems which can be formulated on a lattice in
flux space.  Furthermore, the validity condition (\ref{eq:45}) relates
an experimental quantity (the cosmological constant) to a microscopic
one (an average charge).  We believe that this relation can be pursued
in this context.

\ack We would like to thank Concepci\'on Orna for carefully reading
this manuscript.  This work has been supported by CICYT (grant
FPA-2006-02315) and DGIID-DGA (grant 2007-E24/2).

\end{document}